\documentclass[a4paper]{article}
\usepackage{amsmath}
\usepackage{graphicx}
\usepackage{wrapfig}
\usepackage{enumitem}
\graphicspath{{./Image//}}
\usepackage{float}
\usepackage{geometry}
\usepackage{tikz}
\usepackage{authblk}
\usepackage{hyperref} 
\usepackage{glossaries} 

\newacronym{ml}{ML}{Machine Learning}
\newacronym{qc}{QC}{Quantum Computing}
\newacronym{qml}{QML}{Quantum Machine Learning}
\newacronym{ai}{AI}{Artificial Intelligence}
\newacronym{gb}{GB}{Giga Bytes}
\newacronym{nasa}{NASA}{National Aeronautics and Space Administration}
\newacronym{fun3d}{FUN3D}{Fully Unstructured Navier-Stokes 3-Dimensional}
\newacronym{tb}{TB}{Tera Bytes}
\newacronym{cpu}{CPU}{Central Processing Unit}
\newacronym{usd}{USD}{United States Dollar}
\newacronym{n}{N}{Number}
\newacronym{turing}{Turing}{Turing Machine}
\newacronym{q}{Q}{Quantum}
\newacronym{ket}{KET}{Ket Vector}
\newacronym{bra}{BRA}{Bra Vector}
\newacronym{born}{Born}{Born’s Rule}
\newacronym{amp}{AMP}{Amplitude}
\newacronym{uf}{\ensuremath{U_f}}{Quantum Oracle} 
\newacronym{gpu}{GPU}{Graphics Processing Unit}
\newacronym{petaflop}{PetaFLOP}{Peta Floating Point Operations Per Second}
\newacronym{lasso}{LASSO}{Least Absolute Shrinkage and Selection Operator}
\newacronym{ridge}{Ridge}{Ridge Regression}
\newacronym{enet}{Elastic-net}{Elastic Net}
\newacronym{logreg}{LR}{Logistic Regression}
\newacronym{knn}{K-NN}{K-Nearest Neighbors}
\newacronym{gan}{GAN}{Generative Adversarial Network}
\newacronym{vae}{VAE}{Variational Autoencoder}
\newacronym{gmm}{GMM}{Gaussian Mixture Model}
\newacronym{hmm}{HMM}{Hidden Markov Model}
\newacronym{lda}{LDA}{Linear Discriminant Analysis}
\newacronym{arma}{ARMA}{Auto Regressive Moving Average}
\newacronym{arima}{ARIMA}{Auto Regressive Integrated Moving Average}
\newacronym{qram}{qRAM}{Quantum Random Access Memory}
\newacronym{vqe}{VQE}{Variational Quantum Eigensolver}
\newacronym{qaoa}{QAOA}{Quantum Approximate Optimization Algorithm}
\newacronym{qea}{QEA}{Quantum Evolutionary Algorithms}
\newacronym{qpso}{QPSO}{Quantum Particle Swarm Optimization}
\newacronym{qga}{QGA}{Quantum Genetic Algorithms}
\newacronym{qpca}{QPCA}{Quantum Principal Component Analysis}
\newacronym{qblas}{qBLAS}{Quantum Basic Linear Algebra Subroutines}
\newacronym{qsvm}{QSVM}{Quantum Support Vector Machine}
\newacronym{qknn}{Qk-NN}{Quantum k-Nearest Neighbors}
\newacronym{qkmeans}{Qk-means}{Quantum k-means}
\newacronym{qnn}{QNN}{Quantum Neural Network}
\newacronym{qdt}{QDT}{Quantum Decision Tree}
\newacronym{hhl}{HHL}{Harrow-Hassidim-Lloyd}
\newacronym{qec}{QEC}{Quantum Error Correction}
\newacronym{vif}{VIF}{Variance Inflation Factor}
\newacronym{qiskit}{Qiskit}{Quantum Information Science Kit}
\newacronym{svm}{SVM}{Support Vector Machine}
\newacronym{pca}{PCA}{Principal Component Analysis}
\newacronym{nisq}{NISQ}{Near-Term Intermediate-Scale Quantum}

\makeglossaries 

\newcommand{\citep}[1]{\cite{#1}}

\title{Quantum-Assisted Simulation: A Framework for Developing Machine Learning Models in Quantum Computing}

\author[1]{Minati Rath\thanks{Corresponding author: Minati Rath}}
\author[2]{Hema Date}

\affil[1]{Department of Analytics and Decision Science, IIM Mumbai, India,}
\affil[2]{Department of Analytics and Decision Science, IIM Mumbai, India}

\begin{document}
	
	\newgeometry{
		left=3cm, 
		right=3cm, 
	}
	
	\maketitle
	
\begin{abstract}
	
	\noindent \gls{ml}\label{ml} models are trained using historical data to classify new, unseen data. However, traditional computing resources often struggle to handle the immense amount of data, commonly known as Big Data, within a reasonable time frame. \gls{qc}\label{qc} provides a novel approach to information processing, offering the potential to process classical data exponentially faster than classical computing through quantum algorithms. By mapping \gls{qml}\label{qml} algorithms into the quantum mechanical domain, we can potentially achieve exponential improvements in data processing speed, reduced resource requirements, and enhanced accuracy and efficiency.
	
	\vspace{5pt}
	\noindent In this article, we delve into both the \gls{qc}\label{qc} and \gls{ml}\label{ml} fields, exploring the interplay of ideas between them, as well as the current capabilities and limitations of hardware. We investigate the history of quantum computing, examine existing \gls{qml}\label{qml} algorithms, and present a simplified procedure for setting up simulations of \gls{qml}\label{qml} algorithms, making it accessible and understandable for readers.
	
	\vspace{5pt}
	\noindent Furthermore, we conduct simulations on a dataset using both traditional machine learning and quantum machine learning approaches. We then compare their respective performances by utilizing a quantum simulator.
	
\end{abstract}

\begin{keywords} \end{keywords}
\section{Introduction}
Data is becoming the driving force behind business growth.  Technologies have evolved and are generating huge volume of data from businesses. While it is easy to transmit data securely and seamlessly over internet, storing and retrieving is simple. However, processing them to get meaningful information is a mammoth of task in terms of processing power.  Automation and internet are the largest contributors in data generation.  Artificial intelligence, machine learning, optimisation and simulation models need voluminous data for training models, requiring state-of-the-art high-performance computing.  The computational efforts required to solve any problem depends on size of data and complexity of the equation sets to solve. Some real time simulations need few giga bytes of data for one instance of execution.  One such physics simulation done by \gls{nasa}\label{nasa} is simulation of retro propulsion of a Mars lander using \gls{nasa}\label{nasa}’s \gls{fun3d}\label{fun3d} computational fluid dynamics library \cite{Montes1}.  150 \gls{tb}\label{tb}s of data is used for visualization of the simulation.\\ 
\\Classical computing systems work on the principles of binary bits to solve problems whereas Quantum computation is an entirely new way of information processing. Qubits have the capability to be in quantum states apart from being in binary states.\\
\\Hence, it has become key area to study how quantum computing as a technology is evolving and how machine learning and artificial intelligence models can benefit from them; along with special attention towards error correction and stability of systems and models.\\
\\ The rest of the article is organised as follows.
\begin{enumerate}
\item Section 2 describes classical computing; it's limitations and bottlenecks.\item In Section 3, developmental history and key features of Quantum computers; along with formulation of quantum algorithms are presented.    
\item Section 4 illustrates different supervised and unsupervised machine learning models.
\item In Section 5, we presented research set up for integrating machine learning models into quantum domain; along with challenges impacting quantum machine learning\gls{qml}\label{qml}models.       
\item Finally, conclusion are drawn in section 6.
\end{enumerate}

\section{Classical Computing}

 \begin{wrapfigure}{r}{0.7\textwidth}
	\begin{center}
		\includegraphics[width=1\linewidth]{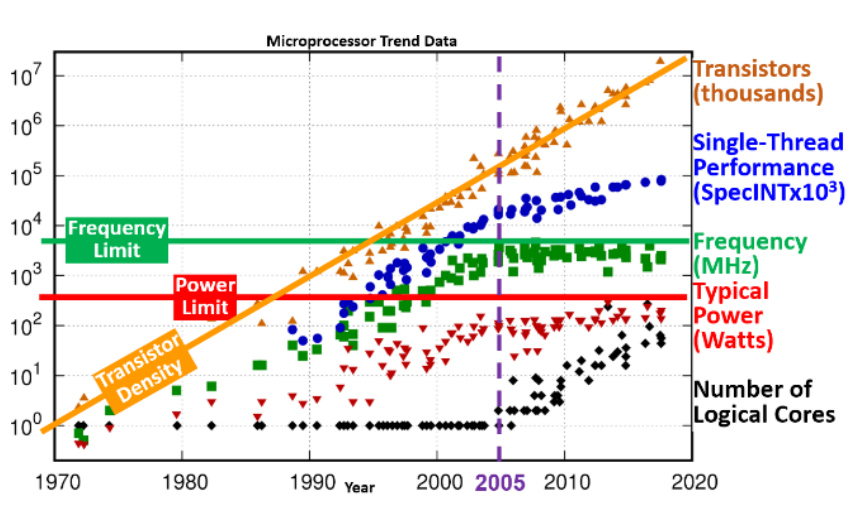}
		\caption{Micro Processor component trend data}
		\label{fig:Micro Processor component trend data}
	\end{center}
\end{wrapfigure} 
Classic computational devices like desktops, laptops, mobile phones and tablets are binary devices whose processing units use two states ie 0s and 1s to store and process information.\\They have finite set of discrete, stable states, with controlled transitions between them.  \\Transistors which are tiny electrical switches are used in computer chips to create classical computing devices.  Advances in new engineering techniques like photonic chips shrunk the number of transistors on a chip making them faster, as photons can travel 20 times faster than electrons and consume less electricity, and operate at a lower cost. 
In 1965, Moore predicted that transistors, capacitors on a chip doubles every two years, yet we pay less for them. He predicted the no of transistors and other components that could fit in a chip, popularly known as “Moore’s Law” \cite{Turcato2018}.\\\\Figure \ref{fig:Micro Processor component trend data} shows growth of transistors, performance, clock speed, power and cores for microprocessor from 1970 till today.\\\\There is exponential growth in the no of transistors compressed in a microprocessor.  Performance too is increasing though not exponentially.  Clock speed, Power consumption and logical cores were reached to the highest limits around 2005. The bottleneck created due to the gap between increase in memory density, processor speed and clock speed resulted lag in data transfer.  To bypass the bottleneck, multiple \gls{cpu}\label{cpu}s were integrated to create parallel processing multi cores.  Some programs and algorithms don't benefit from parallelization.   Such algorithms when subjected to large amount of data fail to perform.\\While classical systems are quite efficient in processing good amount of information, they fail to process very large amount of data.  One such example is calculation of factorial of astronomically large prime numbers.  
	
\section{Quantum Computing}

Quantum computers utilize quantum bits, or qubits, to perform computations. Unlike classical bits, which can exist only in a state of 0 or 1, qubits can simultaneously represent both states due to a phenomenon known as superposition \cite{Nielsen2000}. This unique property allows qubits to perform complex calculations more efficiently than traditional bits. Furthermore, the information stored in qubits possesses intrinsic security features, making it impossible to replicate. Consequently, qubits ensure a high level of privacy for the information they contain.  \\\\	
Quantum properties such as entanglement and teleportation enhance both the security and speed of data transfer. However, it is important to note that quantum states are destroyed upon measurement; once measured, these states collapse into classical bits, represented as either 0 or 1. Therefore, the results of these measurements are recorded in classical registers \cite{Nielsen2000}. 
 
\subsection{History of Quantum Computing}

In 1982, Richard Feynman noted that quantum physical systems cannot be effectively simulated on classical computers, as representing the results of quantum computations with classical systems becomes increasingly infeasible as the system size grows. The computational resources required for such simulations are proportional to the volume of space and time of the quantum system involved \cite{Feynman1982}. 

Building on this foundation, David Deutsch introduced universal quantum circuits in 1985, demonstrating that quantum parallelism could perform certain probabilistic tasks more efficiently than classical computing \cite{Deutsch1985}. 

In 1994, Peter Shor designed the first efficient algorithm capable of factoring extremely large integers into their prime components in polynomial time, a task that is difficult for classical computers. This factorization is crucial for many cryptographic hashing and security algorithms \cite{Shor2002}. 

Lov Grover, in 1997, developed an ultra-fast quantum search algorithm that can search unindexed databases with a complexity of \( O(\sqrt{N}) \) \cite{Grover1996}. 

Furthermore, David P. DiVincenzo proposed five essential requirements for implementing quantum computers \cite{DiVincenzo2000}:

\begin{enumerate}[itemsep=0em]
	\item Scalable and well-characterized qubits serve as the basic building blocks for quantum circuits.
	\item The state of the qubits must be initialized to the ground state \( |00...\rangle \).
	\item A universal set of quantum gates should be established.
	\item Gates must support quantum superposition states that persist longer than the duration of gate operations.
	\item The system must possess qubit-specific measurement capabilities after operations. 
\end{enumerate}

For quantum communication, two additional criteria must be fulfilled:

\begin{enumerate}[itemsep=0em]
	\item Stationary qubits should be able to transition to flying qubits, and vice versa.
	\item Flying qubits must be capable of being transmitted over long distances, which is essential for quantum key distribution.
\end{enumerate}

In 2019, Arute et al. designed a quantum processor featuring 53 qubits, which executed one instance of a quantum circuit a million times in about 200 seconds. In contrast, this computation would take approximately 10,000 years on a classical supercomputer \cite{Arute2019}. 

Qubits are not physical elements or devices; rather, they are logical constructs that can be realized in various systems exhibiting quantum behavior. Quantum computing requires entanglement among all qubits in a system to achieve exponentially faster growth \cite{Vogel2011}. However, qubits are incredibly sensitive to environmental interference. 

Although universal quantum computers remain a theoretical concept, operational quantum computers with a limited number of qubits exist in research laboratories under controlled conditions. Even a small number of qubits can significantly enhance processing capacity, as each additional qubit exponentially increases this capability. Nevertheless, these systems are prone to high error rates due to their sensitivity to environmental factors. Quantum computers that are restricted to specific domain problems under controlled settings are becoming increasingly feasible. Despite their potential to revolutionize computing, the future of quantum computing remains uncertain.

\subsection{Qubits}

Quantum computers perform computations using qubits, or quantum bits. Unlike classical bits, which can be in one of two distinct states (0 or 1), qubits can exist in three states: 0, 1, or a superposition of both. This superposition allows a qubit to represent 0 and 1 simultaneously, a phenomenon that distinguishes quantum computing from classical computing.\\
\noindent Qubits operate on the principle of the "Observation Theory," where the state of a quantum particle is influenced by observation. When observed, the particle behaves differently than when it is not observed, leading to a change in its state. This is akin to the flipping of a coin: while the coin spins in the air, it exists in both the heads and tails states at the same time. Only when it lands does it adopt one of the two definite states. Similarly, unobserved qubits are often considered to be in a state of “spinning,” and can be measured as "up," "down," or "both." \\

\noindent In quantum computing, the standard Dirac notation is commonly used to represent states, which includes the \gls{bra}\label{bra} and \gls{ket}\label{ket} vectors. The \gls{ket}\label{ket} vector represents a state as a column vector, while the \gls{bra}\label{bra} vector is the conjugate transpose of the \gls{ket}\label{ket} vector.

\noindent The fundamental \gls{ket}\label{ket} vectors are defined as follows: $|0\rangle = \begin{bmatrix} 1 \\ 0 \end{bmatrix} \quad \text{and} \quad |1\rangle = \begin{bmatrix} 0 \\ 1 \end{bmatrix}. $

\noindent The corresponding \gls{bra}\label{bra} vectors, \(\langle 0|\) and \(\langle 1|\), are defined as the conjugate transposes of the \gls{ket}\label{ket} vectors:$ \langle 0| = \begin{bmatrix} 1 & 0 \end{bmatrix} \quad \text{and} \quad \langle 1| = \begin{bmatrix} 0 & 1 \end{bmatrix}. $

\subsection{Quantum State}

\begin{wrapfigure}{r}{0.5\textwidth}
	\begin{center}
		\includegraphics[width=7cm,height=6cm]{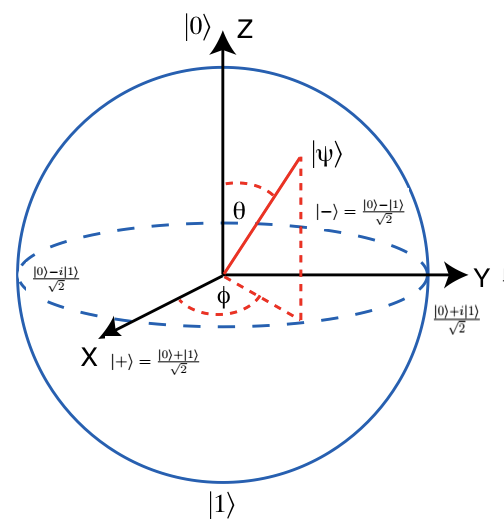}
	\end{center}
	\caption{Bloch Sphere}
	\label{fig:Blotch Sphere}
\end{wrapfigure}

Figure \ref{fig:Blotch Sphere} represents a Bloch sphere. A qubit can be visualized as a unit vector in a 2-dimensional complex vector space. In \gls{ket}\label{ket} notation (where $|\phi\rangle$ denotes a ket, the Dirac notation for vectors), a quantum state is expressed as a column vector with complex coefficients. It is a linear combination of the basic vectors \(|0\rangle\) and \(|1\rangle\).

\noindent Column vectors in $\mathsf{\Xi}$ kets are defined as: $ |\psi\rangle = \alpha|0\rangle + \beta|1\rangle = \begin{bmatrix} \alpha \\ \beta \end{bmatrix} $

\noindent Each quantum state has a well-defined dual vector called a \gls{bra}\label{bra}, which can be obtained by taking the complex conjugate and the transpose of the \gls{ket}\label{ket} vector \cite{Zhou2020}.\\ The dual vectors $\mathsf{\Xi}$ bras are defined as:\\ $ \langle\psi| = [\alpha^*, \beta^*] $\\

\noindent Inner products are calculated by multiplying the \gls{bra}\label{bra} for one state by the \gls{ket}\label{ket} of another, forming a \gls{bra}\label{bra}-ket. The inner products $\mathsf{\Xi}$ brackets are given by:\\ $ \langle\psi|\phi\rangle = [\alpha^*, \beta^*] \begin{bmatrix} \gamma \\ \delta \end{bmatrix} = \alpha^*\gamma + \beta^*\delta $ \\

\noindent When a quantum state measurement is performed to determine whether the state is 0 or 1, a random result is obtained. The probability of measuring a state to be 0 is given by the squared magnitude of its 0 coefficient:\\\\
\noindent $ |0\rangle + |1\rangle \text{ measurement yields } |0\rangle \text{ with probability } |\langle\psi|0\rangle|^2 = |\alpha|^2 $ and \( |1\rangle \text{ with probability } |\langle\psi|1\rangle|^2 = |\beta|^2. \)

\noindent The squared magnitudes must sum to 1 since they represent probabilities. This is known as \gls{born}\label{born}'s rule, and the constraint is referred to as normalization:

$ \langle\psi|\psi\rangle = [\alpha^*, \beta^*] \begin{bmatrix} \alpha \\ \beta \end{bmatrix} = |\alpha|^2 + |\beta|^2 = 1 $\\

\noindent We can calculate the coefficients required to express the same state in the plus-minus basis. States can be expressed in different bases: $\alpha|0\rangle + \beta|1\rangle = (\alpha + \beta)$\\

\noindent The quantum state of a qubit vector \([\alpha, \beta]\), representing a two-level system, is a linear combination of two basic vectors called 0 and 1 (superposition of zero and one states), described by: $|\psi\rangle = \alpha|0\rangle + \beta|1\rangle$, where the coefficients \(\alpha\) and \(\beta\) are known as probability amplitudes. These are complex numbers that satisfy the normalization condition:$|\alpha|^2 + |\beta|^2 = 1$

The basis states are defined as:$|0\rangle = \begin{bmatrix} 1 \\ 0 \end{bmatrix} \quad \text{and} \quad |1\rangle = \begin{bmatrix} 0 \\ 1 \end{bmatrix}\\ $For \(N\) qubits, they can store \(2^N\) binary numbers simultaneously, suggesting massive parallelism. For \(N=2\):$|\psi\rangle = C_0|00\rangle + C_1|01\rangle + C_2|10\rangle + C_3|11\rangle$.  In general, the state can be expressed as:$|\psi\rangle = \sum_{i=0}^{2^n -1} C_i |b_{i,n-1} b_{i,n-2} \ldots b_{i,0}\rangle$

\subsection{Superposition}

\noindent Qubits exhibit a remarkable property known as superposition, allowing them to exist in multiple states at once. This capability enables qubits to store and process significantly more information than classical bits. Additionally, superposition ensures that qubits cannot be replicated or copied, thereby safeguarding the confidentiality of stored information \citep{romero2010toward}. However, this unique feature can pose challenges, particularly when replication is needed for error correction. Superposition forms the basis for various quantum phenomena, including entanglement and quantum teleportation, which utilize the distinctive properties of qubits for advanced computational and communication tasks.

\noindent When a measurement is made or when a qubit interacts with its environment, its superposition collapses to a definitive state, a process often referred to as the "collapse of the wavefunction." The result of the measurement corresponds to one of the potential states in the superposition, with the probabilities of these states determined by the coefficients of their linear combination.

\noindent Superposition also plays a crucial role in quantum interference, a phenomenon arising from the overlap of different quantum states, leading to amplification or cancellation of probabilities. Quantum algorithms frequently exploit interference to achieve accurate results by enhancing the likelihood of the correct outcome while diminishing that of incorrect ones. For instance, Grover’s algorithm utilizes quantum interference to amplify the probability of the correct solution while simultaneously lowering the likelihood of all other possible outcomes. Without superposition, such interference—and consequently the computational advantages of quantum algorithms—would not be achievable.

\noindent In the realm of quantum simulations, superposition is particularly vital. Quantum systems, especially those involving complex molecules and subatomic particles, display behaviors that are difficult for classical computers to replicate. By placing qubits in superposition, quantum computers can naturally simulate the probabilistic nature of these quantum systems, facilitating more efficient simulations of physical systems such as molecules in quantum chemistry and the behaviors of quantum materials.

\noindent In quantum communication and information theory, superposition contributes to the no-signaling theorem, which asserts that neither quantum entanglement nor superposition can be employed to transmit information faster than light. Although entangled qubits can share information instantaneously, measuring the superposition state of one qubit does not directly convey information about the other without classical communication. This characteristic upholds the principle of causality, despite the non-local correlations enabled by quantum entanglement.

\noindent Practically speaking, maintaining superposition is challenging due to the inherent fragility of quantum states. As previously noted, decoherence disrupts the delicate nature of superposition when qubits interact with their surroundings. This presents a significant obstacle in the physical implementation of quantum computers, necessitating advanced techniques such as quantum error correction and fault-tolerant quantum computing to preserve superposition over time. Even minor disturbances from thermal noise or electromagnetic interference can precipitate the collapse of a superposed state, thereby diminishing the effectiveness of quantum computations.

\noindent The concepts of quantum supremacy or quantum advantage are fundamentally tied to the exploitation of superposition. Quantum algorithms that utilize superposition can solve specific problems exponentially faster than their classical counterparts. For example, random circuit sampling—a task used to illustrate quantum advantage by Google’s Sycamore processor—benefits from the capacity of quantum systems to maintain and manipulate exponentially large superpositions, a feat that classical computers struggle to achieve efficiently.

\subsection{Entanglement}

\noindent Quantum entanglement is a fundamental characteristic exclusive to quantum systems, absent in classical systems. In the process of entanglement, two or more particles interact in such a way that their individual states become inseparable, even when they are physically separated by a significant distance. Consequently, measurements performed on entangled qubits at one end will yield correlated results at the other end. Entangled qubits demonstrate a high level of coordination, making them suitable for tasks that require synchronization or coordination \citep{4483}.

\noindent Combinatorial optimization problems typically possess multiple solutions, each with an equal probability of occurrence. When these problems are given to binary devices, they need to individually solve each possibility. In contrast, quantum computers leverage the unique characteristic of entanglement to access multiple problem solutions simultaneously. Quantum computers do not solve problems through parallel computation, where all possible solutions are attempted simultaneously to reduce execution time. Instead, qubits, capable of existing in two states simultaneously, explore multiple solutions while in superposition, and upon measurement, produce the optimal result \citep{Farhi2014}.

\noindent Quantum entanglement refers to a unique and exclusive connection between two qubits, wherein no other qubit elsewhere can access or share this entangled information. This characteristic provides quantum networks with enhanced privacy and security capabilities. In quantum communication systems, information is transmitted through entangled qubits between physically separated quantum processors. This remarkable phenomenon is often referred to as teleportation. During the teleportation process, the sender does not require knowledge of the recipient's location or the specific details of the quantum state being teleported \citep{Bouwmeester_1997}.

\subsection{Guideline for Quantum Algorithms Framework}

\noindent Typically, a hybrid quantum-classical algorithm consists of two main components: a quantum part and a classical part. The quantum part of the algorithm involves generating quantum parametrized circuits that prepare quantum states to process data inputs and perform calculations.

\noindent The following are the detailed steps involved in this process.

\begin{enumerate}[itemsep=0em]
	\item Determine the required number of qubits for modeling the problem, and subsequently, set their initial state to the ground state $|0\rangle$. 
	\item Create uniform superposition of all possible states of the qubits. 
	\item Generate a uniform superposition by preparing the qubits in a state that encompasses all possible states with equal probability. 
	\item Encode the input data into the various superposition states that have been generated.
	\item Sequentially apply unitary and multi-qubit gates to induce the necessary phase shifts required for implementing quantum algorithms or a \gls{uf}\label{uf}. A Quantum Oracle can be described as a black box that allows the reuse of other circuits or algorithms. 
	\item The computation result is typically stored in one or a few of the qubit states.
	\item To extract the result, the qubits are disentangled, and a measurement is performed. 
	\item To increase the probability of obtaining the correct unit, the amplitudes of other states are reduced through amplification, ensuring that measuring the final state will yield the desired unit with a high degree of certainty.
	
\end{enumerate}

\noindent In the classical component, the outcomes of the qubit measurements are recorded and utilized to optimize parameters through classical registers.

\noindent The following steps are involved:
\begin{enumerate}[itemsep=0em]
	\item Perform measurements on one or multiple qubits and store in classical registers. 
	\item Optimize the parameters obtained from qubits. It involves fine-tuning the values obtained from qubits to achieve optimal performance or desired results. This optimization process typically includes iteratively adjusting these parameters based on specific objectives or criteria.
	\item The optimized results from classical registers are encoded back to the qubits. The process is repeated until the final result is derived.
	\item By performing simulations of the entire quantum-classical hybrid circuit multiple times, typically in the range of a few thousand iterations, one can obtain more precise probability counts. This iterative approach not only enhances the fidelity of the algorithm but also improves its overall performance.
	\item Implement error mitigation techniques, such as error correction codes or error mitigation algorithms to enhance the accuracy and reliability of the algorithm.
\end{enumerate}

\section{Machine learning Models}
Machine learning models have evolved using data as their backbone. The goal is to identify patterns, data classification, trends in large datasets, and many others. These models are majorly categorized into supervised, unsupervised, and reinforcement learning \cite{Oladipupo2010}. \\
\noindent Data is used to train algorithms to optimize model parameters to recognize the training data. Hence, data is divided into training and test datasets. The model learns better with a large amount of training data. Accuracy is determined using test data.  There has been an exponential growth in the amount of training data used to train \gls{ml}\label{ml} and AI models in the last seven years. \gls{ml}\label{ml} models are consuming \gls{petaflop}\label{petaflop} of data in a day. It is a mammoth task for \gls{cpu}\label{cpu}s and \gls{gpu}\label{gpu}s to process this data to extract meaningful outcomes.

\noindent Regression, classification, and clustering are the most common machine learning tasks \cite{Oladipupo2010}\cite{Rong2018}.

\subsection{Supervised machine learning}
	
Supervised machine learning uses labelled dataset to generate and train algorithms. The model is used to predict outcome or classify an unknown input. It's a model defined with set of parameters.
\[ 	y=h(x|\theta)+ \varepsilon \]
where h(x$|\theta)$ is model, $\theta$ is set of parameters and $\varepsilon$ is error in model.\\ y is numeric when h(x$|\theta)$ is Regression function.\\ y is class code (e.g 0 or 1) when h(x$|\theta)$ is a Discriminant function.\\
\\Regression tasks are supervised models.  They are predictive models characterized by labelled datasets that have continuous, numeric and categorical target variables \cite{Oladipupo2010}\cite{Doan}.  New inputs are passed to the model to predict the output. 
Linear regression is most basic algorithm that uses a straight line between single dependent variable and target variable. The target variable is continuous variable. $ y=mx+c+\epsilon$, where m is slope,c is intercept and $\epsilon$ is model error.
\\Corresponding hypothesis is represented by $ h_0(x)=\theta_0+\theta_1+\epsilon $, where $\theta_0$ is constant and $\theta_1$ is regression coefficient. The idea is to choose $\theta_0$, $\theta_1$; so that $h_0(x)$ is close to y for training set (x,y).\\
Error is computed by actual value minus predicted value for all the rows in the dataset. So, for given $x^{(i)}$, actual output is $y^{(i)}$ and regression line is $(\theta_0+\theta_1x^{(i)})$.\\
\[ \begin{split}
Error (\epsilon) &=((\theta_0+\theta_1x^{(i)})-y^{(i)}))\\
&=(h_0(x^{(i)}-y^{(i)}))
\end{split} \]
The objective is to reduce the overall error for the training dataset.  The objective function/cost function is sum of squared error.\\\\
Cost Function : $ J(\theta_0,\theta_1)=\frac{1}{2m}\sum_{i=1}^m (h_\theta(x^{(i)})-y^{(i)})^2 $, where m is number of data points.\\\\
Squared Error cost function : $ Cost(h\theta(x^{(i)},y^{(i)})=\frac{1}{2}(h\theta(x^{(i)})-y^{(i)})^2$.  When the no of dependent variables is two or more, it attempts to fit a straight hyperplane to the dataset.  These models are susceptible to outliers and performs poorly when there are non-linear relationships\cite{Rong2018}.  
Polynomial regression can be used to improve the efficiency to certain extent.  Here, the regression line is in the form of a curve. Quadratic and cubic polynomial regression are commonly used. 
Fitting a quadratic function: $ \theta_0+\theta_1x+\theta_2x^2 +\epsilon $
Higher degree of polynomials can be added to models. $ \theta_0+\theta_1x+\theta_2x^2+\theta_3x^3+\theta_4x^4+\epsilon$
It can be tricky, time consuming and may introduce over fitting to the training data. Both over fitting, which performs well with training data has high variance and low bias, under fitting wherein the model does not perform well neither with training data nor generalizes test data, are not preferable in any model. Balance between high variance and high bias will result in a better model.  To overcome under fitting or high bias, new parameters are added to model to increase model complexity, and thus reducing high bias.  Regularization and reduced model complexity overcome under fitting.\\   
Regularized linear regression models like \gls{lasso}\label{lasso}, \gls{ridge}\label{ridge} and \gls{enet}\label{enet} outperform simple regression models \cite{Oladipupo2010}.  Regularization keeps same no of features and penalizes large coefficients to avoid .  \\
\gls{logreg}\label{logreg} is a supervised classification task for categorical target variables.  It predicts a class instead of a real numeric value. $ [y\in {0,1}] $, 0 is negative class and 1 is positive class. Hypothesis for \gls{logreg}\label{logreg}  is represented by 
\[\begin{split}
& 0 \leq h_\theta(x) \leq 1 \\
& h_\theta(x) = g(\theta^Tx)\\
& g(Z)=\frac{1}{1+e^{-z}}
\end{split}
\] 
$h_\theta(x)$ is estimated probability and g(z) is sigmoid function.
It uses class probabilities and logistic function to predict a class. It works well when classes are linearly separable.
Logistic Regression Cost function:
\[ J(\theta_0,\theta_1)=\frac{1}{m}\sum_{i=1}^m Cost(h_\theta(x^{(i)}),y^{(i)}) \]
\[ Cost(h_\theta(x),y)=
\begin{cases}
-log(h_\theta(x)),&\textbf{if } y=1\\
-log(1-h_\theta(-x))& \textbf{if } y=0
\end{cases} \]
y is always 0 or 1.\\
\\
Over fitting issues can be handled either by reducing number of features or reducing the magnitude of parameters keeping all the features.\\
Regression models with lot of features can be improved by regularisation ie penalizing the higher coefficients. The model parameters are regularized with a tunable penalty strength.  \\\\
Regularisation for linear regression: $ j(\theta)=\frac{1}{2m}\Bigl[\sum_{i=1}^m (h_\theta(x^{(i)})-y^{(i)})^2+\lambda\sum_{j=1}^n \theta_j^2 \Bigr] $, $\lambda$ is regularisation parameter. It helps fitting the training data well by keeping the value of parameters small. A large value of $\lambda$ results in under fitting model.  L1(\gls{lasso}\label{lasso}) regularisation adds absolute value of magnitude of parameters as penalty term.  L2(\gls{ridge}\label{ridge}) regression adds squared magnitude of parameters as penalty term. 
\\
Both linear and logistic regression models can be updated easily using stochastic gradient descent but are not flexible to perform well with unknown as well as complex data. \\Regression uses R-Square as another evaluation metric, whose value is between 0 and 1. 0 indicates that the independent variables in the model does not explain variability in the dependent variable and 1 indicates that it contributes fully to explain variability in the dependent variable.  It can be negative for really bad model.  When new independent variables are added to model, either R2 increases or remains constant.  R2 never decreases by adding variables to model. The disadvantage is, we can't conclude that increasing complexity of our model makes it more accurate.  The Adjusted R2 is improved form of R2. It is used to compare goodness of fit for the model. The adjusted R-Square only increases if addition of new predictor/s improves the model accuracy.
\subsection{Supervised Machine Learning}

Supervised machine learning utilizes labeled datasets to generate and train algorithms. The model predicts outcomes or classifies unknown inputs, defined by a set of parameters: $ y = h(x | \theta) + \varepsilon $, where \( h(x | \theta) \) is the model, \( \theta \) is the set of parameters, and \( \varepsilon \) is the error in the model. Here, \( y \) is numeric when \( h(x | \theta) \) represents a regression function, and \( y \) is a class code (e.g., 0 or 1) when \( h(x | \theta) \) is a discriminant function.

\noindent Regression tasks are characterized by labeled datasets that contain continuous, numeric, and categorical target variables \cite{Oladipupo2010}\cite{Doan}. New inputs are passed to the model to predict the output. 

\noindent Linear regression is the most basic algorithm, which uses a straight line to model the relationship between a single dependent variable and a target variable: $ y = mx + c + \epsilon $, where \( m \) is the slope, \( c \) is the intercept, and \( \epsilon \) is the model error. The corresponding hypothesis is represented by: $ h_0(x) = \theta_0 + \theta_1 + \epsilon $, Here, \( \theta_0 \) is a constant and \( \theta_1 \) is the regression coefficient. The objective is to choose \( \theta_0 \) and \( \theta_1 \) such that \( h_0(x) \) closely approximates \( y \) for the training set \((x,y)\). \noindent The error is computed as the difference between the actual value and the predicted value for all rows in the dataset: \[\begin{split}
	\text{Error} (\epsilon) &= (h_0(x^{(i)}) - y^{(i)}) \\
	&= (h_0(x^{(i)}) - y^{(i)})
\end{split}\]
\noindent The goal is to minimize the overall error for the training dataset. The objective function, or cost function, is the sum of squared errors:\[
J(\theta_0, \theta_1) = \frac{1}{2m} \sum_{i=1}^m (h_\theta(x^{(i)}) - y^{(i)})^2\], where \( m \) is the number of data points. The squared error cost function can be expressed as:\[
\text{Cost}(h_\theta(x^{(i)}), y^{(i)}) = \frac{1}{2}(h_\theta(x^{(i)}) - y^{(i)})^2
\]
\noindent When the number of dependent variables is two or more, the model attempts to fit a straight hyperplane to the dataset. However, these models are susceptible to outliers and perform poorly in the presence of non-linear relationships \cite{Rong2018}. 

\noindent Polynomial regression can improve efficiency to some extent by allowing the regression line to take the form of a curve. Quadratic and cubic polynomial regressions are commonly used, where fitting a quadratic function can be expressed as: $ \theta_0 + \theta_1 x + \theta_2 x^2 + \epsilon $

\noindent Higher-degree polynomials can also be added to the model: $\theta_0 + \theta_1 x + \theta_2 x^2 + \theta_3 x^3 + \theta_4 x^4 + \epsilon $

\noindent However, this can lead to overfitting, where the model performs well on training data but poorly on unseen data. Both overfitting (high variance, low bias) and underfitting (poor performance on both training and test data) should be avoided. Striking a balance between high variance and high bias results in a better model. To mitigate underfitting or high bias, new parameters can be introduced to increase model complexity, while regularization techniques can reduce model complexity.

\noindent Regularized linear regression models, such as Lasso, Ridge, and Elastic Net, outperform simple regression models \cite{Oladipupo2010}. Regularization maintains the same number of features while penalizing large coefficients to avoid overfitting. \\

\noindent Logistic regression (\gls{logreg}\label{logreg}) is a supervised classification task for categorical target variables, predicting a class instead of a numeric value:$ y \in \{0, 1\} $, where 0 represents the negative class and 1 represents the positive class. The hypothesis for logistic regression is represented by:
\[
\begin{split}
	0 \leq h_\theta(x) \leq 1 \\
	h_\theta(x) = g(\theta^T x) \\
	g(Z) = \frac{1}{1 + e^{-z}}
\end{split}
\] Here, \( h_\theta(x) \) is the estimated probability, and \( g(z) \) is the sigmoid function. Logistic regression uses class probabilities and the logistic function to predict a class and works well when classes are linearly separable.\\

\noindent The logistic regression cost function is given by: $ J(\theta_0, \theta_1) = \frac{1}{m} \sum_{i=1}^m \text{Cost}(h_\theta(x^{(i)}), y^{(i)}) $\\

\noindent The cost is defined as: $ \text{Cost}(h_\theta(x), y) =
\begin{cases}
	-\log(h_\theta(x)), & \text{if } y = 1 \\
	-\log(1 - h_\theta(x)), & \text{if } y = 0
\end{cases} $ Here, \( y \) is always 0 or 1.\\

\noindent Overfitting can be addressed by reducing the number of features or by controlling the magnitude of parameters while retaining all features. Regression models with a large number of features can be improved through regularization, penalizing higher coefficients. \\\\

\noindent The regularization for linear regression is defined as: \[ j(\theta) = \frac{1}{2m} \left[\sum_{i=1}^m (h_\theta(x^{(i)}) - y^{(i)})^2 + \lambda \sum_{j=1}^n \theta_j^2 \right] \]\noindent  where \( \lambda \) is the regularization parameter, helping fit the training data well while keeping the parameter values small. A large value of \( \lambda \) can lead to underfitting. L1 regularization (Lasso) adds the absolute value of the magnitude of parameters as a penalty term, while L2 regularization (Ridge) adds the squared magnitude of parameters as a penalty term.\\

\noindent Both linear and logistic regression models can be easily updated using stochastic gradient descent but may struggle to perform well with complex or unknown data. \\

\noindent Regression utilizes R-squared as an evaluation metric, with values ranging from 0 to 1. A value of 0 indicates that the independent variables do not explain variability in the dependent variable, while 1 indicates complete explanatory power. It can be negative for poorly performing models. Adding new independent variables to the model may either increase or keep R-squared constant, but it never decreases. However, the increasing complexity of a model does not guarantee greater accuracy. The Adjusted R-squared is an improved metric, used to compare the goodness of fit for models. The Adjusted R-squared only increases if the addition of new predictors enhances model accuracy.

\subsection{Unsupervised Machine Learning}

Clustering is a form of lazy unsupervised learning, effectively assigning classes to each vector from a set of multivariate data inputs without the intent of treating new inputs. The inputs are grouped into clusters, with data members in a cluster being similar to each other and dissimilar to members in other clusters. Lazy learning algorithms do not have a training phase; that is, generalization of the training data occurs in advance, and the model does not learn a discriminative function but rather memorizes the training dataset. An example of this is the k-Nearest Neighbors (k-NN) algorithm. Clustering can serve as an effective data reduction technique for analysis \cite{Rong2018}. \\

\noindent With the tremendous amount of data collected from internet applications, the majority of it is unlabeled and exists in various forms, such as images, text, and speech. The focus of unsupervised learning is to model the underlying structure of the data, searching for hidden patterns without predetermined outputs.\\\\ Clustering algorithms are categorized into two types:

\begin{itemize}
	\item[] Hierarchical clustering algorithms, which build a hierarchy of clusters either in a bottom-up (agglomerative) or top-down (divisive) fashion.
	\item[] Partitioning clustering algorithms, which divide the dataset into distinct clusters without hierarchical relationships.
\end{itemize}

\noindent The two most common clustering algorithms are:

\begin{itemize}
	\item[] k-Means clustering, which requires the number of clusters (k) to be predefined, works by minimizing the sum of squared distances from each point to its cluster center.
	\item[] k-Medoids clustering, similar to k-Means, but uses the actual data points (medoids) as centers of clusters instead of the mean.
\end{itemize}

\noindent A clustering objective can be represented as:  \[
\begin{split}
	\text{min } J &= \sum_{k=1}^K \sum_{x_i \in C_k} || x_i - \mu_k ||^2 \\
	\text{where } \mu_k &= \frac{1}{|C_k|} \sum_{x_i \in C_k} x_i
\end{split}
\]

Here, \( J \) is the total cost function for clustering, \( K \) is the number of clusters, \( C_k \) is the set of points assigned to cluster \( k \), and \( \mu_k \) is the center of cluster \( k \). Clustering can be used for customer segmentation, grouping similar items, and other applications, as it helps uncover natural groupings in data without pre-defined labels \cite{Oladipupo2010}.

\noindent Clustering can be further expanded by combining both supervised and unsupervised learning techniques for semi-supervised learning, where both labeled and unlabeled data are used to create training sets.

\subsection{Reinforcement Learning}

Reinforcement learning focuses on learning through interactions with an environment, optimizing actions to maximize cumulative rewards. It employs a trial-and-error approach, where an agent learns to take actions that yield the highest reward in a specific context. Each action taken by the agent can result in different outcomes, influencing future decisions. \\

\noindent Reinforcement learning can be broken down into key components, including the agent, the environment, actions, states, and rewards. The agent interacts with the environment by taking actions and observing the resulting state and reward. The goal is to learn a policy that defines the best action to take in a given state to maximize the expected cumulative reward. The process is defined by the Markov decision process (MDP):\[
\begin{split}
	\text{State } s_t & \text{ at time } t \\
	\text{Action } a_t & \text{ taken at state } s_t \\
	\text{Reward } r_t & \text{ received after taking action } a_t
\end{split}
\]

\noindent The cumulative reward can be expressed as: $ R_t = r_t + \gamma r_{t+1} + \gamma^2 r_{t+2} + \ldots $, where \( \gamma \) is the discount factor, determining the present value of future rewards. Reinforcement learning models can optimize complex tasks, including robotic control, game playing, and resource allocation.

\noindent  In conclusion, machine learning models offer a diverse set of techniques and methodologies, each suited to different types of problems. Understanding the strengths and weaknesses of each model type is crucial for developing effective solutions in various domains.

\section{Quantum Information Processing and Machine Learning}

Classical computational devices—such as desktops, laptops, mobile phones, and tablets—are binary systems that utilize two states (0s and 1s) to store and process information. These devices operate with a finite set of discrete, stable states and controlled transitions between them. While they are efficient for processing a significant amount of information, they struggle with extremely large datasets. A notable example of this limitation is the calculation of the factorial of astronomically large prime numbers \citep{Shor2002}.

\noindent In contrast, quantum computers perform computations using quantum bits, or qubits. Unlike classical bits, qubits can exist in three states: 0, 1, or both simultaneously. This capability arises from a phenomenon known as superposition, allowing qubits to represent a combination of states. Moreover, qubits possess unique properties that make them non-replicable, ensuring a high level of privacy for any information stored within them. Quantum features such as entanglement and teleportation further enhance the security and speed of data transfer. However, it is important to note that quantum states are destroyed upon measurement; once measured, they collapse to classical bits (0 or 1), and these measurements are subsequently stored in classical registers \citep{vogel2011quantum}.

\noindent Machine learning algorithms can significantly benefit from the advantages of quantum computing. These algorithms need to be redesigned to operate on quantum computers, enabling the development of quantum machine learning algorithms that can tackle complex problems beyond the current capabilities of classical approaches \citep{biamonte2017quantum}.

\noindent One major distinction between classical and quantum systems is that classical systems are deterministic, while quantum systems are probabilistic. The primary objective is to map NP-hard algorithms and their subroutines into quantum information theory, creating quantum circuits and subroutines that can run efficiently on quantum computers.

\subsection{The Research Setup for Quantum Machine Learning}

The objective of this research setup is to establish a framework for modeling machine learning and optimization problems within linear quantum physical systems. The specific goals include:

\begin{itemize}
	\item[] To leverage the properties of quantum computing to simulate big data and address computationally challenging (NP-hard) problems.
	\item[] To model non-linear functions that can be executed on linear quantum mechanical systems.
	\item[] To explore available quantum libraries for designing and implementing quantum circuits.
	\item[] To create quantum machine learning and optimization models utilizing digital quantum systems, incorporating appropriate unitary and multi-qubit quantum gates, as well as analog quantum systems based on the time evolution of qubits.
	\item[] To encode classical data points as parameters in quantum superposition states for subsequent processing.
	\item[] To perform matrix operations on these encoded parameters to execute various algorithmic steps.
	\item[] To observe qubits and measure the outputs of these operations, followed by post-processing to interpret the results. These results will be passed to classical registers for cost function calculation, with subsequent updates computed for the parameters.
	\item[] To iteratively run the circuit with updated parameters until the cost function is minimized.
	\item[] To extract averages from measurement results with higher probabilities for each qubit or output qubit.
	\item[] To compare results obtained from both classical and quantum approaches.
	\item[] To conduct an analysis of the Classical-Quantum variational hybrid model and analog adiabatic model, assessing their applicability in various contexts.
	\item[] To identify machine learning algorithms suitable for addressing specific quantum computing problems.
	\item[] To model complex combinatorial optimization problems as Hamiltonians, representing the total energy of a system, and to minimize this energy to identify the ground state, which corresponds to the optimal solution of the problem.
	\item[] To identify intractable machine learning and optimization steps and adapt them to the quantum domain.
	\item[] To pinpoint applications that can benefit from small numbers of noisy qubits in Noisy Intermediate-Scale Quantum (NISQ) systems.
\end{itemize}

\subsection{Research Progress and Opportunities in QML}

Presently, Noisy Intermediate-Scale Quantum (\gls{nisq}\label{nisq}) computers typically possess a range of 5 to 5000 qubits. Although these systems exhibit sufficient computational capabilities to achieve quantum supremacy, the accuracy of their outputs cannot be reliably verified due to the presence of noise and errors \cite{Arute2019}\cite{Boixo2018}. These quantum systems can sustain their quantum state for brief periods, typically on the order of $10^{-3}$ seconds. Once measured, qubits lose their quantum state, and classical computers are employed to store the measurement results. Universal quantum simulators can generate qubit quantum states that closely approximate physical states. By leveraging these quantum states and algorithms, computational space and time costs can be minimized.   \\
\noindent QML employs \gls{qram}\label{qram} (Quantum Random Access Memory) for storing quantum states. \gls{qram}\label{qram}, comprising qubits, consists of three main components: 1. A memory array that addresses the quantum superposition of memory cells, 2. An input register, and 3. An output register \cite{Giovannetti2008}. \\
\noindent \gls{qram}\label{qram} serves as a storage and retrieval system for classical information. It combines classical memory with a quantum addressing scheme, enabling the access and retrieval of specific elements within the memory \cite{Blank}. \\
\noindent Machine learning problems are tackled using hybrid quantum algorithms such as the Variational \gls{vqe}\label{vqe} \cite{Liu}. \\ \noindent The \gls{vqe}\label{vqe} involves providing a set of parameters to a Quantum Processing Unit (QPU), which in turn produces a set of measurements. These measurements are then utilized by a classical computer to calculate the energy value using averaging methods \citep{Liu}\citep{McClean_2016}. \\ \noindent \gls{qaoa}\label{qaoa} (Quantum Approximate Optimization Algorithm) is used in combinatorial optimization problems. Optimization problems try to find the optimal solution to a given problem in finite time from a set of infinite possible solutions \cite{McClean_2016} \cite{Choi}\citep{Guerreschi2019}. Optimization problems encompass a range of objectives, including minimizing costs, distances, and consumption, and maximizing profits or outputs. It is possible to convert a maximization problem into a minimization problem and vice versa by appropriately transforming the objective function.  \\ \noindent \gls{qea}\label{qea} refers to a class of evolutionary algorithms inspired by quantum principles. By leveraging quantum systems, \gls{qea}\label{qea} accelerates the evolutionary process by simultaneously evaluating multiple solutions in parallel. This parallel evaluation enhances the probability of discovering the optimal solution \citep{Wang2018}. Within this framework, algorithms such as Quantum Particle Swarm Optimization (\gls{qpso}\label{qpso}) and Quantum Genetic Algorithms (\gls{qga}\label{qga}) are applied to solve diverse problems like job shop scheduling, flow shop scheduling, procurement, feature selection, and vehicle routing problems \citep{Singh201636}\citep{Zouache201826}\citep{Qin20161}\citep{Beheshti2015402}. \\ \noindent \gls{qpca}\label{qpca} (Quantum Principal Component Analysis), a dimensionality reduction technique, utilizes \gls{qram}\label{qram} and quantum tomography to construct the eigenvectors of the principal components characterized by their significant eigenvalues \citep{Zhou2020}\citep{Zhou2020}\citep{Lloyd2014}. Quantum tomography refers to the procedure of validating a quantum state. It involves extracting the eigenvalues of the qubit gates while accounting for noise. Additionally, it quantifies the fidelity of the gates through measurement \citep{Lloyd2013}\citep{Mohseni_2008}. \\
\noindent Linear systems of equations can be solved using subroutines such as \gls{hhl}\label{hll} (Harrow-Hassidim-Lloyd), quantum phase estimation, and Hamiltonian simulations. These subroutines are specifically designed for addressing the challenges associated with solving linear equations in a quantum computing context \citep{Harrow2009}.  \\
\noindent In the context of calculating distances between objects, subroutines like K-nearest neighbors employ a quantum approach. In this method, unclassified data is stored in one quantum register, while the training set is stored in another register. An ancillary qubit is utilized in a third register. The resulting calculations are then stored in the first register \citep{Schuld2017}.  \\
\noindent \gls{qblas}\label{qblas} (Quantum Basic Linear Algebra) refers to the quantum counterparts of traditional linear algebra operations. These quantum operations are utilized in various Quantum Machine Learning (QML) algorithms, including \gls{qknn}\label{qknn}, \gls{qsvm}\label{qsvm}, \gls{qknn}\label{qknn}means, \gls{qnn}\label{qnn}, and \gls{qdt}\label{qdt} \citep{Schuld2017}. \\
\noindent In the context of machine learning, kernels are derived from inner products of features mapped to a feature map. In quantum circuits, classical data is encoded to calculate kernels by computing dot products of features. These computed kernels are then fed into \gls{svm}\label{svm}s for further analysis and processing \citep{Lloyd2013}. \\

Recent advancements in QML research include:

\begin{itemize}
	\item[] \text{Quantum Federated Learning:} This approach focuses on training models across decentralized data sources without exchanging the data itself. By combining the strengths of quantum computing and federated learning, researchers aim to improve privacy and efficiency in machine learning tasks \citep{Basso2021}.
	
	\item[] \text{Quantum Neural Networks:} Recent studies have explored the potential of quantum neural networks (QNNs) to outperform classical counterparts in specific tasks, demonstrating advantages in training speed and capacity for handling complex datasets \citep{Dunjko2016}.
	
	\item[] \textbf{Quantum-enhanced Reinforcement Learning:} Researchers are investigating the application of quantum algorithms in reinforcement learning frameworks to achieve faster convergence rates and enhanced exploration capabilities in complex environments \citep{Zhang2021}.
	
	\item[] \text{Quantum Feature Mapping:} New techniques for mapping classical data to quantum states have been developed to enhance the performance of quantum algorithms, particularly in high-dimensional datasets \citep{Havlíček2019}.
	
	\item[] \text{Quantum Support Vector Machines:} The performance of quantum support vector machines (QSVMs) has been compared with classical SVMs, showing promising results in classification tasks, particularly in cases involving large datasets with complex patterns \citep{Huang2021}.
	
	\item[] \text{Hybrid Quantum-Classical Approaches:} Ongoing research focuses on integrating classical and quantum machine learning models, enabling the development of more robust algorithms that leverage the strengths of both paradigms \citep{Cai2020}.
\end{itemize}

These advancements underscore the potential for Quantum Machine Learning to revolutionize various fields by enabling more efficient and powerful data analysis methods.

\subsection{Challenges with Quantum \gls{ml} Models\label{ml}}

Both the fields of machine learning and quantum computing are characterized by their fragmented and diverse nature. Researchers are actively investigating the potential for synergy between these two domains, aiming to leverage the unique capabilities of quantum computing in machine learning applications. However, the integration of quantum computing and machine learning poses several challenges that need to be addressed. These challenges can significantly impact the development and performance of Quantum Machine Learning (QML) models. 

\noindent Current quantum computers are not capable of outperforming classical models due to limited qubit counts. Qubits experience noise and decoherence, leading to the loss of information and errors as they interact with the surrounding environment. In quantum algorithms, a series of gates are applied linearly to qubit registers; however, the presence of decoherence imposes limitations on the number of operations that a quantum circuit can effectively perform. This represents a significant hardware challenge in quantum computing, requiring the development of quantum hardware with a large number of fully interconnected and high-fidelity qubits. Achieving fault-tolerant universal systems while maintaining qubit quality is a crucial objective in overcoming this challenge.

\noindent \gls{nisq} systems undergo effective isolation and cooling processes to mitigate noise and errors. \gls{qec} techniques are employed in \gls{nisq} algorithms to minimize the impact of decoherence. \gls{qec} aims to correct quantum qubits using a reduced number of auxiliary ancilla qubits while preserving the superposition states and coherence time. However, improving the accuracy of algorithms remains a challenge. Advancements in error correction schemes hold the potential to reduce the number of auxiliary qubits required for correcting a single qubit, thereby enhancing overall algorithm accuracy.

\noindent The limited connectivity of qubits poses a challenge in designing efficient circuits for complex algorithms. While theoretical analysis may indicate the required number of qubits for a calculation, physical implementation necessitates the inclusion of auxiliary qubits for error correction and additional processing qubits to account for circuit depth and complexity. Shallow algorithms, such as the variational quantum eigensolver, require fewer qubits in comparison. Accurate initialization of qubits to their desired starting state is crucial, as errors introduced during initialization propagate and accumulate through subsequent gate operations, leading to significant computational errors.

\noindent The coherence time of qubits should exceed the duration required for gate operations. As more gate operations are performed on the qubits, it becomes necessary to increase the qubits' ability to retain quantum information for a longer duration.

\noindent Translating machine learning problems into the quantum domain does not have foolproof methods yet. It requires the development of strategies and techniques to effectively map and configure machine learning problems onto quantum systems. Some of the key challenges include:

\begin{enumerate}
	\item[] Optimal number of qubits and circuit depth to perform \gls{ml} algorithms.
	\item[] Choosing and configuring the optimizer.
	\item[] Designing effective feature maps with fewer parameters to optimally encode classical data.
	\item[] Minimizing the loss function.
	\item[] Designing optimizers tailored specifically for quantum loss functions.
	\item[] Addressing hard and intractable \gls{ml} algorithms, which are often solved heuristically. Quantum optimization algorithms can superposition all possible search states to evaluate cost functions.
	\item[] Extending the coherence time for superpositioned qubits to reduce the number of qubits needed for algorithms.
	\item[] Developing robust methods for the initialization of qubits to minimize propagation of errors during gate operations.
	\item[] Addressing the challenge of scalability, as larger quantum systems are required to handle more complex datasets and algorithms.
	\item[] Enhancing the interpretability of quantum models to understand and trust the outcomes produced by QML algorithms.
	\item[] Bridging the gap between theoretical research and practical implementations to enable the application of QML in real-world scenarios.
\end{enumerate}

\section{Experimental Results}

\begin{figure}[H]
	\begin{center}
		\includegraphics[width=0.5\linewidth]{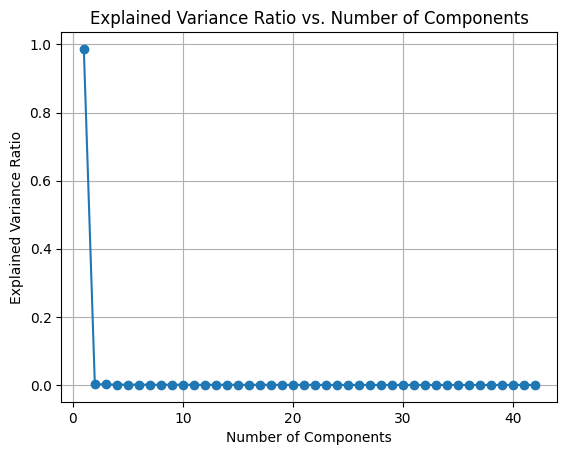}
		\caption{Explained Variance ration with Elbow point}
		\label{fig:ExplainedVarianceElbow}
	\end{center}
\end{figure}

We experimented with a machine learning algorithm and the \gls{qiskit}\label{qiskit}, quantum simulator to tackle a customer churn classification problem within the telecommunications domain. We explored, tested, or tried out different approaches using these tools to address the problem.
\\The dataset comprises 20 features, including a binary target column indicating churn, and encompasses a total of 7,043 distinct customers. This dataset can be found on the Kaggle website.\\
Among the features, the following are categorical: gender, SeniorCitizen, Partner, Dependents, PhoneService, MultipleLines, InternetService, OnlineSecurity, OnlineBackup, DeviceProtection, TechSupport, StreamingTV, StreamingMovies, Contract, PaperlessBilling, PaymentMethod, and Churn. On the other hand, tenure, MonthlyCharges, and TotalCharges are numeric features.
We observed a correlation coefficient of 0.83 between Tenure and TotalCharges, leading us to exclude TotalCharges from our analysis. Fortunately, no outliers or missing values were found in the dataset.\\\\
Regarding the PhoneService feature, we noticed a multicollinearity \gls{vif}\label{vif} score above 12, prompting us to eliminate it from further processing. After reevaluating the \gls{vif}\label{vif}, we discovered that MonthlyCharges had a score close to 6, and thus we dropped it as well. Additionally, we decided to remove the customerID feature as it does not contribute to the model.
\\\\
Out of the 16 features, all except 'tenure' have undergone one-hot encoding, resulting in a total of 42 columns. The classes in the churn label are significantly imbalanced, with a total of 1,869 instances belonging to the minority 'Yes' class and 5,174 instances belonging to the majority 'No' class. Due to the limitations of the simulator in handling large volumes of data, we opted to retain the minority classes and performed under sampling on the majority classes, resulting in a total of 3,738 records.
\\\\
We attempted to determine the optimal number of components where the explained variance ratio diminishes significantly. By plotting the number of components against the explained variance ratio, we identified an elbow point at 23. Figure \ref{fig:ExplainedVarianceElbow} plots explained variance Vs Number of components.  \\\\
We have opted for the Support Vector Machine (\gls{svm}\label{svm}) algorithm to classify churn data. We conducted experiments using both classical and hybrid quantum-classical approaches with the \gls{qiskit}\label{qiskit} BasicAir simulator. The BasicAir simulator allows a maximum of 24 qubits, so we reduced the dimensionality of our dataset to 2, 10 and 15 vector dimensions using \gls{pca}\label{pca}. The table \ref{tab:PCA Variance} below shows the training and testing scores, with an 80:20 train-test split ratio.  We could not train all 23 dimensions due to limited simulator capabilities.

\begin{table}[H]
	\centering
	\caption{\gls{pca} and their Explained Variance}
	\hspace{4cm}
	\label{tab:PCA Variance}
	\begin{tabular}{|p{1.5cm}|p{10cm}|}
		\hline
		\textbf{Vector Dimensions} &  \textbf{Explained Variance} \\
		\hline
		\gls{pca} =2 & [0.9876, 0.0029] \\
		\hline
		\gls{pca}=10 & [9.8763, 2.8920, 1.2026, 9.6864,
		8.3936, 8.2999, 5.9614, 5.8387,5.2604, 4.8225] \\
		\hline			
		\gls{pca}=15 & [[9.870, 3.1892, 1.2432, 1.0133,
		8.8052, 8.3666, 6.2122, 5.9178,
		5.4195, 4.9875, 4.4907, 4.1575,
		4.0606, 3.9272, 3.7539 \\
		\hline				
		\gls{pca}=20 & [9.8711, 2.9649, 1.2579, 1.0080,
		8.8054, 8.7051, 6.2967, 6.1114,
		5.6773, 4.9847, 4.5460, 4.2820,
		4.0724, 4.0206, 3.8747, 3.3482,
		3.1868, 2.9462, 2.1622, 1.9710]\\
		\hline		
	\end{tabular}
\end{table}

\vspace{20pt} 

\begin{quote}
	\centering
	Elbow Point Index: 23\\
	Explained Variance Ratio at Elbow Point: 8.126234391114803e-33 \\
	Cumulative Explained Variance at Elbow Point: 1.0000000000000002 \\
\end{quote}

\vspace{2pt} 

\begin{table}[H]
	\centering
	\caption{Train and test accuracy from quantum simulator and classical system}
	\hspace{6cm}
	\label{tab:Quantum Classical Results}
	\begin{tabular}{|c|c|c|c|}
		\hline
		\textbf{} & \textbf{\gls{pca}\label{pca}=2} & \textbf{\gls{pca}\label{pca}=10} & \textbf{\gls{pca}\label{pca}=15} \\
		\hline\hline
		\textbf{Train Score: Quantum} & 0.5986 & 0.9817 & 0.9809  \\
		\hline
		\textbf{Test Score: Quantum} & 0.6002 & 0.617 & 0.5748   \\
		\hline \hline
		\textbf{Train Score: Classical} & 0.7137 & 0.7444 & 0.7461  \\
		\hline
		\textbf{Train Score: Classical} & 0.7366 &  0.7647 & 0.7473  \\
		\hline
		
	\end{tabular}
\end{table}
\hspace{4cm}
  
The results of train and test accuracies are listed in table \ref{tab:Quantum Classical Results}. As a result of applying under sampling techniques on our data and the additional dimensionality reduction through \gls{pca}\label{pca}, the train and test accuracies obtained from the quantum system exhibit significant bias. Similarly, the classical system also demonstrates considerably low train and test accuracies.

\section{Conclusion}
	
Quantum computing, machine learning, and evolutionary algorithms are currently popular areas of research. In our presentation, we provided an overview of the historical development of quantum computing and machine learning, as well as their integration to achieve quantum advantage. We discussed the steps involved in designing quantum algorithms and the experimental setup required for Quantum Machine Learning models. Additionally, we highlighted some of the main challenges associated with current hardware in implementing \gls{qml}\label{qml} models on Near-Term Intermediate-Scale Quantum (\gls{nisq}\label{nisq}) systems. Quantum information processing, utilizing the principles of quantum mechanics such as superposition, entanglement, the no-cloning theorem, and quantum annealing, presents exciting possibilities for future advancements in artificial intelligence, machine learning, deep learning, and combinatorial optimization models.   

\printglossary[type=\acronymtype]

\section*{Acknowledgements}
The authors have no acknowledgements to declare.

\section*{Author Declarations}

\subsection*{Conflict of Interest}
The authors have no conflicts to disclose.

\subsection*{Ethical Approval and Consent to Participate}
No ethical approval or consent to participate was required for this study.

\subsection*{Consent for Publication}
All authors consent to the publication of this manuscript.

\subsection*{Availability of Supporting Data}
The data supporting the findings of this study are openly available in the Kaggle repository: \url{https://www.kaggle.com/datasets/blastchar/telco-customer-churn}

\subsection*{Competing Interests}
The authors declare that they have no competing interests.

\subsection*{Authors' Contributions}
All authors contributed equally to this work.

\subsection*{Funding}
This research received no specific grant from any funding agency in the public, commercial, or not-for-profit sectors.

\bibliographystyle{unsrt}
\bibliography{LR}

\section*{Author information}
Authors and Affiliations \\

Phd of Analytics and Decision Science, IIM Mumbai, India. 

Minati Rath.

Professor of Analytics and Decision Science, IIM Mumbai, India. 

Hema Date.

\section*{Corresponding authors}

Correspondence to \href{minati06@gmail.com}{Minati Rath} or  \href{hemadate@iimmumbai.ac.in}{Hema Date}.

\end{document}